\newcommand{\tindex}[1]{{\scriptstyle {\rm#1}}}

\documentclass[12pt]{article}

\usepackage{natbib}
\usepackage{epsfig}
\usepackage{graphicx}

\usepackage[]{epic,eepic,epsf}

\begin{document}

\renewcommand{\refname}{\normalsize \bf \em References}

\pagestyle{myheadings}
\markright{
  Phase Transitions {\bf 75}, 257 (2002)
  }
\title{\bf INTERFACE MOTION IN DISORDERED FERROMAGNETS}
\author{
L.\ ROTERS\footnote{Email: lars@thp.uni-duisburg.de},
S.\ L\"UBECK AND K. D. USADEL
\\*[0.2cm]
  {\it \small Theoretische Tieftemperaturphysik,} \\
  {\it \small Gerhard-Mercator-Universit\"at Duisburg,
   47048 Duisburg, Germany}
}
\maketitle

\begin{abstract}
%
  We consider numerically the depinning transition in the
  random-field Ising model. Our analysis reveals that the three and
  four dimensional model displays a simple scaling behavior whereas
  the five dimensional scaling behavior is affected by logarithmic
  corrections. This suggests that $d=5$ is the upper critical
  dimension of the depinning transition in the random-field Ising model.
  Furthermore, we investigate the so-called creep regime (small
  driving fields and temperatures) where the interface velocity is
  given by an Arrhenius law. 
%
\end{abstract}


\section*{1. INTRODUCTION} \label{intro}

\begin{figure}[t] 
  \begin{center}
  \includegraphics[height=7cm]{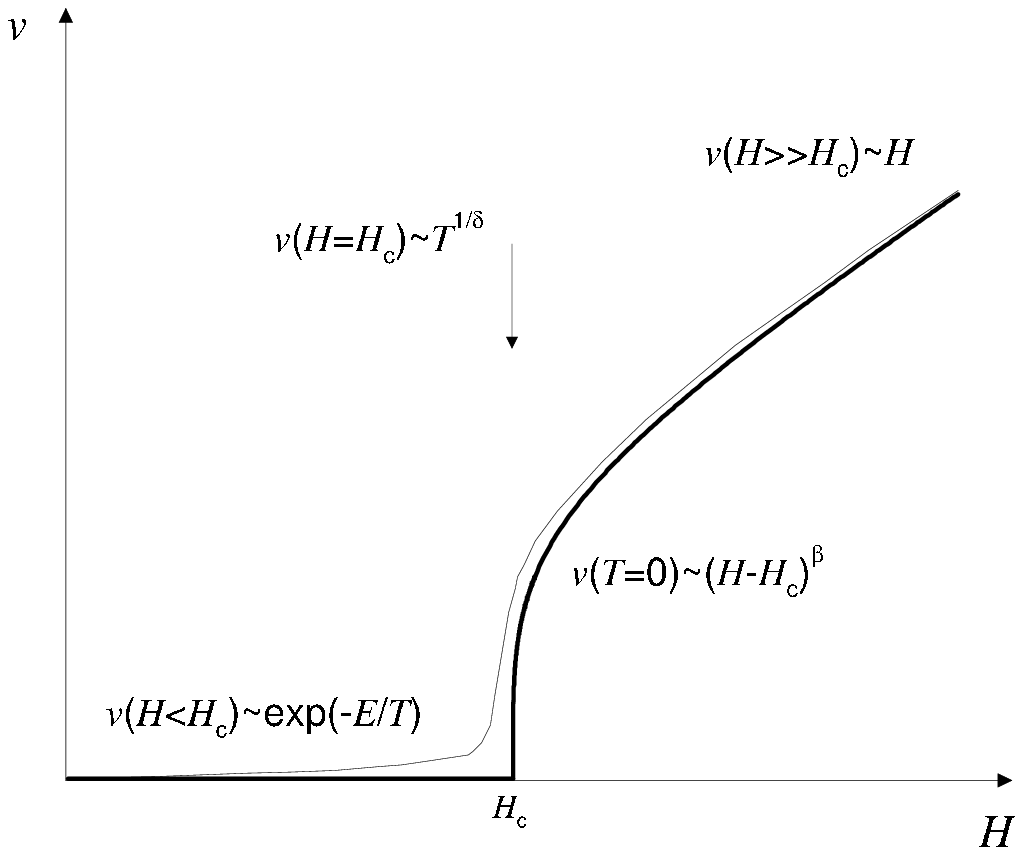}
    \caption{
        Sketch of the interface velocity $v$ on the driving field $H$ and 
        temperature $T$.
        The bold line corresponds to $T=0$.
        With increasing temperature the depinning transition is
        smeared out (thin solid line).
        In the creep regime the interface velocity obeys an Arrhenius
        law. 
        }
    \label{fig:general}
  \end{center}
\end{figure}

Using the random-field Ising model (RFIM) we have studied the dynamics of
driven interfaces in disordered ferromagnetic systems by means of
Monte Carlo simulations. The interface separates regions of opposite
magnetization and it is driven by a homogeneous field $H$.
Without thermal fluctuations ($T=0$) its dynamics is affected by the
disorder in the way that the interface moves only if a sufficiently
large driving field is applied. With decreasing driving field the
interface velocity vanishes at a critical threshold $H_\tindex{c}$ and
is pinned below $H_\tindex{c}$, due to the random-field. The field
dependence of the interface velocity is given by
\begin{equation}
  v ( h ) \sim h^\beta
  \label{eq:d3}
\end{equation}
with $h=(H-H_\tindex{c})/H_\tindex{c} \geq0$, i.e.\ this
pinning/depinning transition can be considered as a continuous phase
transition. A sketch of this behavior is shown in Fig.\
\ref{fig:general}.

The paradigm of the depinning transition is the Edwards-Wilkinson
equation \citep*{Edwards-82} with quenched disorder (QEW).
The critical exponent $\beta$ of the QEW equation has been estimated
within an renormalization group approach \citep*{Nattermann-92}
It is expected that the critical behavior of both the RFIM and the QEW
equation is characterized by the same exponents, i.e.\ both models are
in the same universality class see e.g.\ \citep*{Bruinsma-84,Amaral-94}.

In this paper we determine numerically the critical exponent $\beta$
of the depinning transition in the RFIM in various dimensions.
We compare our results with those of the renormalization group
approach mentioned above. In particular, we consider the five
dimensional case where logarithmic corrections to the above scaling
ansatz, Eq.\ (\ref{eq:d3}), occur. These logarithmic corrections are a
typical feature of the upper critical dimension where the so-called
mean field solution becomes valid.

For finite temperatures no pinning of the interface occurs due to the
thermal fluctuations. Even for driving fields well below
$H_\tindex{c}$ the interface is expected to exhibit a so-called creep
motion, see Fig.\ \ref{fig:general}, in which the interface velocity
obeys an Arrhenius law. This Arrhenius law has been found for the QEW
equation \citep*{Chauve-98} and we compare these results with those of
the RFIM.

\section*{2. THE RFIM MODEL} \label{model}

In order to study the dynamics of moving interfaces numerically, we
have investigated the $d$-dimensional RFIM which is characterized by
the Hamiltonian,
\begin{equation}
  {\cal H} =
  -\frac{J}{2}\, \sum_{\langle i,j \rangle} S_i \, S_j
  -H \, \sum_{i} S_i
  - \sum_{i} h_i\,S_i .
\end{equation}
The sum in the first term is taken over all pairs of neighbored spins.
It describes the exchange interaction between neighboring spins
($S_i=\pm 1$) and a parallel alignment of nearest neighbors is
energetically favored. Additionally, the spins are coupled to a
homogeneous driving field $H$ and to a quenched local random-field
$h_i$ with $\langle h_i h_j \rangle \propto \delta_{ij}$ and
$\langle h_i \rangle = 0$. The probability density $p$ that the
local random-field takes some value $h_i$ is given by
\begin{equation}
  p(h_i) = 
  \left\{
    \begin{array}{ccl}
      (2\Delta)^{-1} & \;{\rm for}\; &|h_i| < \Delta\\
      0 & & \mbox{otherwise}  ,
    \end{array}
  \right.
\end{equation}
i.e.\ $h_i$ is uniformly distributed. Throughout the paper we consider
the case $\Delta=1.7$. Using both simple cubic (sc) and body centered
cubic (bcc) lattices with antiperiodic boundary conditions we started
each simulation for an initially flat interface. In non-disordered
systems ($\Delta=0$) the interface moves for any finite driving field. 
This limiting behavior for $\Delta \to 0$ can be recovered if
the interface moves along the diagonal direction of simple cubic (sc)
lattices or along the $\hat{z}$-direction of body centered cubic (bcc)
lattices, for details see \citep*{Nowak-98}.

In our Monte Carlo simulations the dynamics of the interface motion is
given by a Glauber dynamics with transition probabilities according to
a heat-bath algorithm, see e.g.\ \citep*{Binder-97}. The interface
velocity is the basic quantity in our investigations. Since a moving
interface corresponds to a magnetization $M$ which increases with time
$t$ (given in Monte Carlo steps per spin) the interface velocity is
obtained according to $v \sim \langle {\rm d}M/{\rm d}t \rangle$, where
$\langle ... \rangle$ denotes an appropriate disorder average.

\section*{3. CRITICAL INTERFACE DYNAMICS} \label{critic}

\begin{figure}[t] 
  \begin{center}
    \includegraphics[height=7cm]{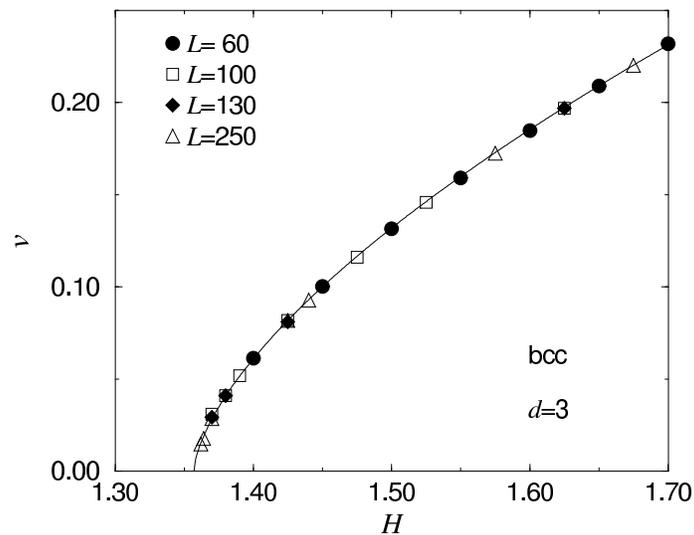}
    \caption{
      Dependence of the interface velocity $v$ on the driving field $H$
      in the three dimensional model for different system sizes $L$.
      The data can be fitted (dashed line) according to
      Eq.\ (\ref{eq:d3}).
      From the fit one obtains
      $H_\tindex{c}=1.357 \pm 0.008$ and $\beta = 0.642 \pm 0.06$. 
      }
    \label{fig:d3}
  \end{center}
\end{figure}

In the following we analyse the interface dynamics in the vicinity of
the critical threshold $H_\tindex{c}$ at zero temperature ($T=0$)
on a bcc lattice. The interface velocity as a function of the driving
field for the three dimensional model is shown in Fig.\ \ref{fig:d3}.
The data suggest that $v$ vanishes continuously at the critical
threshold $H_\tindex{c}$. Fitting the data to the above ansatz,
Eq.\ (\ref{eq:d3}), one obtains $\beta=0.642 \pm 0.06$ and
$H_\tindex{c}=1.357\pm 0.008$. This agrees with previous results.
In a simple cubic lattice an exponent $\beta=0.6 \pm 0.11$ was
observed for an interface moving along the $[001]$ direction 
\citep*{Amaral-94}, and $\beta=0.66 \pm 0.04$ was observed for an
interface moving along the $[111]$ direction \citep{Roters-99},
respectively. Additionally, these values agree within the error-bars
with the value $\beta = 2/3$ found for the QEW equation with
quenched disorder~\citep*{Nattermann-92}.

\begin{figure}[t] 
  \begin{center}
   \includegraphics[height=7cm]{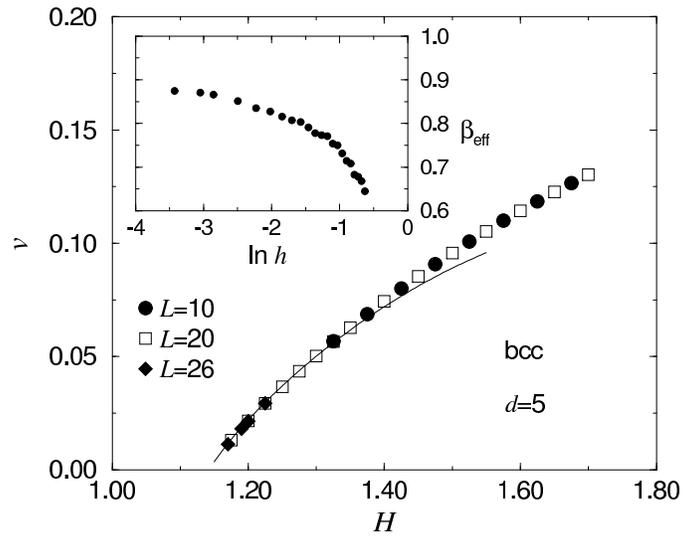} 
   \caption{
     Dependence of the interface velocity $v$ on the driving field $H$
     in the five dimensional model for different system sizes $L$.
     The dashed line is a fit according to Eq.\ (\ref{eq:d5})
     where $\beta_\tindex{MF}=1$ has been used.
     As can be seen from the data, the fit works quite well for
     $H \to H_\tindex{c}$ and
     one obtains $H_\tindex{c}=1.1425 \pm 0.001$ and
     $x_h=0.42 \pm 0.1$.
     The inset shows the logarithmic derivative
     $\beta_\tindex{eff}=\partial \ln v / \partial \ln h$.
     The strong curvature indicates that the velocity field dependence
     cannot be described by Eq.\ (\ref{eq:d3}).
     }
   \label{fig:d5}
 \end{center}
\end{figure}

The above scaling behavior of the interface velocity, Eq.\
(\ref{eq:d3}), is also found in two \citep*{Nowak-98} and four
dimensions (not shown). But in five dimensions the simple scaling
ansatz Eq.\ (\ref{eq:d3}) does not describe the velocity-field
dependence in a convincing manner. In the inset of Fig.\ \ref{fig:d5}
we show the logarithmic derivative
\begin{math}
  \beta_\tindex{eff}=\partial \ln v / \partial \ln h
\end{math}
for various driving fields. Instead of a saturation for $h\to0$ the
effective exponent $\beta_\tindex{eff}$ displays a strong curvature
which contradicts the scaling ansatz in Eq.\ (\ref{eq:d3}).

In the following we will show that the data for $d=5$ are consistent
with the assumption that the scaling behavior is given by
Eq.\ (\ref{eq:d3}) modified by logarithmic corrections. Usually
logarithmic corrections affect the scaling behavior at the upper
critical dimension $d_\tindex{c}$. Below $d_\tindex{c}$ the critical
exponents depend on the dimension of the system while for
$d > d_\tindex{c}$ they are independent of the particular dimension
and are given by those of the mean field theory. At $d_\tindex{c}$
itself the mean field scaling behavior is modified by logarithmic
corrections. The above mentioned renormalization group approach to the
QEW equation predicts the values $d_\tindex{c}=5$ and
$\beta_\tindex{MF}=1$. 

Inspired by the usual scaling behavior at the critical dimension
obtained from renormalization group approaches, see e.g.\
\citep*{brezin-76}, we assume that the asymptotic $(h \to 0)$ scaling
behavior of the velocity is given in leading order by
\begin{equation}
  v \sim |h|^{\beta_\tindex{MF}}
  \left| \ln h \right|^{x_h}
  \, \mbox{.}
  \label{eq:d5}
\end{equation}
with $\beta_\tindex{MF}=1$ and some unknown exponent $x_h$.  
The corresponding data as well as the fit according to
Eq.\ (\ref{eq:d5}) are shown in Fig.\ \ref{fig:d5}. From the fit which
works for small values of $h$ we obtain $H_\tindex{c}=1.143 \pm 0.01$ and
$x_h=0.42\pm0.1$. It remains to be shown that the scaling behavior above
$d_\tindex{c}$ is given by the pure mean field scaling behavior.
Simulations of the six dimensional model are in progress and the
results will be published elsewhere.

\section*{4. THE CREEP MOTION} \label{creep}

\begin{figure}[t] 
  \begin{center}
   \includegraphics[height=7cm]{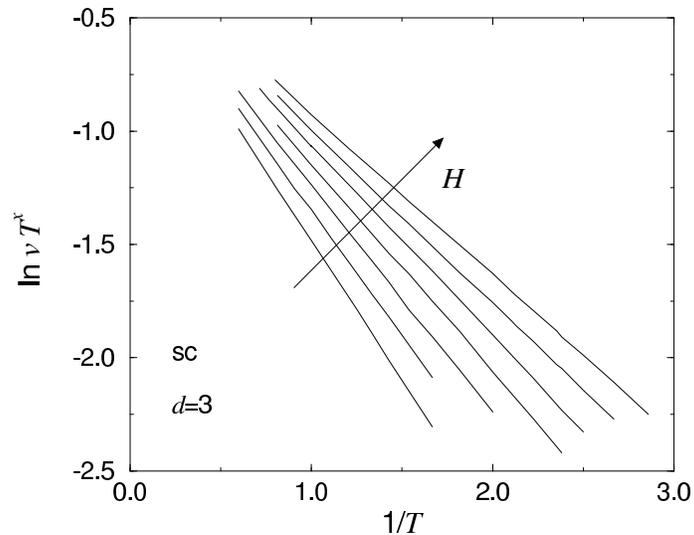} 
   \caption{
     Interface velocities obtained at different driving fields
     $H =0.3, 0.35, 0.4, ..., 0.6$ and rescaled according to 
     Eq.\ (\ref{eq:arr}) and (\ref{eq:pre}).
     Varying $x$ we obtain nearly straight lines for
     $x=0.79 \pm 0.09$. 
     }
   \label{fig:lines}
 \end{center}
\end{figure}

\begin{figure}[t] 
  \begin{center}
   \includegraphics[height=7cm]{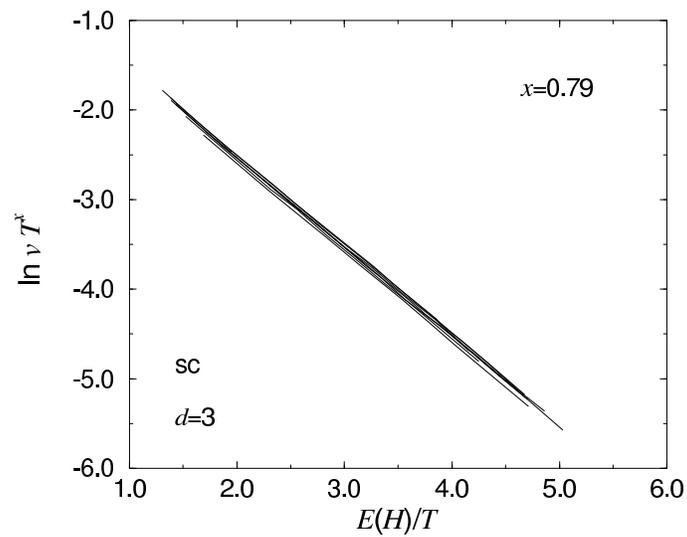}
   \caption{
     Rescaled interface velocities.
     The data coincide with those shown in Fig.\ \ref{fig:lines}.
     They are rescaled with the numerically determined energy
     barrier $E(H)$.
     The coalescence of the data suggests 
     $c(H) \approx {\rm const}$. 
     }
   \label{fig:coal}
 \end{center}
\end{figure}

In the previous section we focused our attention to the critical
behavior at zero temperature. Including thermal fluctuations no
pinning of the interface takes place, see Fig.\ \ref{fig:general}.
For example, at $H=H_\tindex{c}$ the velocity grows with the
temperature according to $v \sim T^{1/\delta}$
\citep{Nowak-98,Roters-99}. The interface moves even for driving
fields well below the critical threshold. In this case the interface
velocity is expected to obey an Arrhenius law 
\begin{equation}
  v \sim C(H,T) \, \exp \left[-\frac{E(H)}{T}\right]
  \label{eq:arr}
\end{equation}
which is characterized by its prefactor $C(H,T)$ and an effective
energy barrier~$E(H)$.  Recently \citep*{Chauve-98} investigated the
creep regime in the QEW equation and found that the prefactor of the
Arrhenius law is given by
\begin{equation}
  C(H,T)=c(H) \, T^{-x}
  \label{eq:pre}
\end{equation}
with $c(H)$ being some function of the driving field. The temperature
dependence is characterized by an exponent $x$. We apply this ansatz
to our numerically obtained interface velocities in the 3$d$ RFIM. 
Figure \ref{fig:lines} shows the data rescaled according to 
Eq.\ (\ref{eq:arr}) and (\ref{eq:pre}). As can be seen from the data,
we obtain nearly straight lines for $x=0.79\pm 0.09$ suggesting that
the temperature dependence of the prefactor in Eq.\ (\ref{eq:arr}) as
well as the Arrhenius law itself are a proper description of the
velocity in the creep regime. The slope of the lines in Fig.\
\ref{fig:lines} corresponds to the effective energy barrier $E(H)$.
Plotting $\ln v\,t^x$~vs.~$E(H)/T$ the data collapse onto a single
curve, see Fig.\ \ref{fig:coal}, without making any assumption about $c(H)$.
This suggests that $c(H) \approx {\rm const}$ holds. 

However, details of our results do not coincide with those of
\citep*{Chauve-98}. In particular, these authors observed a significant
field dependence of the prefactor $c(H) \sim H^\mu$ with $\mu>0$ and
found a negative exponent $x$. Especially the opposite sign of $x$
is remarkable.

\section*{5. SUMMARY}

We investigated the dynamics of driven interfaces for the random-field
Ising model (RFIM). In our analyses we focus on the field and temperature
dependence of the interface velocity and we consider the depinning
transition especially at temperature $T=0$. In the three dimensional
model we find that $v(H)$ obeys a power law characterized by an
exponent $\beta$. In $d=5$ the scaling behavior can be described by
the mean field exponent modified by logarithmic corrections.
This suggests that the upper critical dimension of the depinning
transition in the RFIM is $d_\tindex{c}=5$. 

We also investigate the interface dynamics in the so-called creep
regime which occurs for driving fields well below the critical
threshold and small but finite temperatures. As predicted by a
renormalization group approach the creep behavior is characterized
by an Arrhenius law. But the details of the Arrhenius law differ from
the predicted behavior.  
\\*[0.2cm]

\noindent{\bf \em Acknowledgements}
\\

\noindent 
This work was supported by the DFG via GK~277
{\it Struktur und Dynamik heterogener Systeme} (University
of Duisburg) and SFB~491
{\it magnetische Heteroschichten: Struktur und elektronischer
  Transport}
(Universities of Duisburg and Bochum).


\newpage

\end{document}